\documentclass[graybox]{svmult}

\usepackage{type1cm}        
\usepackage{amsmath}
\usepackage{makeidx}         
\usepackage{graphicx}        
\usepackage{multicol}        
\usepackage[bottom]{footmisc}
\definecolor{ourcolor}{rgb}{0.7, 0.25, 0.05}

\makeindex             

\begin{document}

\title*{{\color{ourcolor}Explaining muon magnetic moment and AMS-02 positron excess in a gauged horizontal symmetric model}}
\author{Gaurav Tomar} 
\institute{Gaurav Tomar \at Physical Research Laboratory, Ahmedabad 380009, India,\\ 
           Indian Institute of Technology, Gandhinagar 382424, India,\\ \email{tomar@prl.res.in}}
%
%
\maketitle
\abstract{ We extended the standard model with a fourth generation of fermions to explain the discrepancy in 
the muon magnetic moment and to describe the positron excess observed by AMS-02 experiment. We introduce a gauged $SU(2)_{HV}$ 
horizontal symmetry between the muon and the 4th generation lepton families and identified the 4th generation right-handed 
neutrino as the dark matter with mass $\sim 700$ GeV. The dark matter annihilates through $SU(2)_{HV}$ gauge boson into final 
states $(\mu^+ \mu^-)$ and $(\nu^c_\mu ~\nu_\mu)$. The $SU(2)_{HV}$ gauge boson with mass $\sim 1.4$ TeV gives the required
contribution to the muon $(g-2)$ and satisfy the experimental constraint from BNL measurement.}
\section{Introduction}
\label{sec:1}
The discrepancy between the experimental measurement \cite{Bennett:2008dy} and the standard model (SM) projection of muon anomalous
magnetic moment (in short muon $g-2$) and the excess of positrons 
observed by AMS-02 \cite{Accardo:2014lma} are the two interesting 
signals which may have a common beyond standard model explanation.\\
In the standard model, the muon anomalous magnetic moment behaves as $a_{\mu} \propto m^2_{\mu}/M^2_{W,Z}$ and its contribution is
$a^{\rm SM}_{\mu}=19.48 \times 10^{-10}$ \cite{Beringer:1900zz}. But SM contribution is $3.6~\sigma$ away from the measured 
\cite{Bennett:2008dy} value of muon $g-2$, which states
\begin{equation}
 \Delta a_{\mu} \equiv a^{\rm Exp}_{\mu}-a^{\rm SM}_{\mu}= (28.7 \pm 8.0) \times 10^{-10},
 \label{er}
\end{equation}
where $a_{\mu}$ is the anomalous magnetic moment in the the unit of $e/2m_{\mu}$.
The mass suppression in muon $g-2$ can be evaded by proposing a horizontal symmetry \cite{Baek:2001kca,Tomar:2014rya}.\\
AMS-02 experiment \cite{Accardo:2014lma} has observed the excess of positron over cosmic-ray background, which goes upto
$\sim 500$ GeV. The dark matter (DM) annihilation with leptonic final states $\mu$ or $\tau$ 
can explain the observed excess very well. The absence of antiproton excess over cosmic-ray background also indicates towards
a leptophilic dark matter.\\
We introduce a 4th generation fermions family and propose a $SU(2)_{HV}$ gauge symmetry between 4th generation leptons and 
muon families. As an artefact of $SU(2)_{HV}$ gauge symmetry, we have new contributions to muon $g-2$ from $SU(2)_{HV}$ 
gauge boson $\theta^+$ and different scalars. In this model, we identified the 4th generation right-handed neutrino 
$\nu_{\mu^\prime R}$ as dark matter. The annihilation of dark matter takes place through $SU(2)_{HV}$  gauge boson $\theta_3$
with only possible final states $(\mu^+ \mu^-)$ and $(\nu^c_\mu~\nu_\mu)$. The dark matter stability is insured by taking
4th generation charged lepton heavier than dark matter. The required cross-section to explain AMS-02 positron excess is
$\sigma v_{\chi\chi\rightarrow\mu^+\mu^-} = 2.33 \times 10^{-25}\rm cm^3/sec$, which is larger than the cross-section 
$\sigma v_{\chi\chi\rightarrow SM} \sim 3\times 10^{-26}\rm cm^3/sec$ required for getting correct relic density \cite{6}.
\section{Model}
\label{sec:2}
Keeping the exact structure of SM, we add the 4th generation of quarks $(c^\prime, s^\prime)$ and 
leptons $(\nu^\prime_\mu, \mu^\prime)$ (of both chiralities) into it. In addition we also introduce three-right handed neutrinos
and extend the standard model gauge group by proposing a $SU(2)_{HV}$ horizontal symmetry. In this model, we have taken $e$ and
$\tau$ families as a singlet of $SU(2)_{HV}$ for simplicity  and explain muon $g-2$ and AMS-02 positron excess simultaneously.\\
The left-handed muon and 4th generation lepton families are denoted by $\Psi_{Li\alpha}$ and their right-handed neutral and 
charged counterparts are denoted by $N_{R\alpha}$ and $E_{R\alpha}$ respectively. We denote the left-handed electron and tau 
doublet by $\psi_{eLi}$ and $\psi_{\tau Li}$ and their right-handed equivalent by $e_R$ and $\tau_R$ respectively. In this
model, the gauge fields correspond to $SU(2)_L\times U(1)_Y\times SU(2)_{HV}$ groups are $A^a_\mu,B_\mu$ and 
$\theta^a_\mu~(a=1,2,3)$ with gauge couplings $g,g^\prime$ and $g_H$ respectively. The gauge couplings of muon
and 4th generation lepton families are given as,
\begin{align}\nonumber
 {\cal L}_{\psi} &=
   i\bar \Psi_{Li\alpha}\gamma^\mu
 \left(\partial_\mu-\frac{i}{2}g\tau \cdot A_\mu+ ig^\prime B_\mu-\frac{i}{2}g_H\tau \cdot \theta_\mu\right)_{ij;\alpha\beta}
 \Psi_{Lj\beta}\\\nonumber
 &+ i\bar E_{R\alpha}\gamma^\mu
 \left(\partial_\mu+ i2 g^\prime B_\mu-\frac{i}{2}g_H\tau \cdot \theta_\mu\right)_{\alpha\beta}E_{R\beta}
 + i\bar N_{R\alpha}\gamma^\mu
 \left(\partial_\mu-\frac{i}{2}g_H\tau \cdot \theta_\mu\right)_{\alpha\beta}N_{R\beta}\\
 \label{eq:gc}
\end{align}
from eq.\ref{eq:gc}, it is clear that the ``neutral current" of $SU(2)_{HV}$ contributes to the 
annihilation process, 
$(\nu_{\mu^\prime}\nu_{\mu^\prime})\rightarrow\theta^*_3\rightarrow (\mu^+\mu^-),(\nu^c_\mu ~\nu_\mu)$, which is appropriate
for AMS-02 and relic density. The ``charged-current" contributes to the muon $g-2$.\\
There exist strong bounds on the 4th generation from the higgs production at LHC, so we extend the higgs sector (in addition
to $\phi_i$) by a scalar $\eta^\beta_{i\alpha}$. As a $SU(2)$ doublet $\eta^\beta_{i\alpha}$ lifts the bounds from higgs 
overproduction and 125 GeV mass eigenstate is mainly constituted by $\eta$. To generate masses for $SU(2)_{HV}$ gauge bosons,
we introduce another scalar $\chi_\alpha$, which is a doublet under $SU(2)_{HV}$. The quantum numbers of the scalars and 
fermions in the model are shown in table.\ref{tab:1}. After corresponding scalars take their vacuum expectation values (vevs), 
the masses of the gauge bosons come,
\begin{equation*}
 M^2_W = \frac{g^2}{2}(2\langle\eta\rangle^2 + \langle\phi\rangle^2),~~M^2_Z = 
          \frac{g^2}{2} {\rm sec}^2\theta_W  (2\langle\eta\rangle^2 + \langle\phi\rangle^2),
 ~~M^2_A = 0,
 \end{equation*}
 \begin{equation}
  M^2_{\theta^+} = g_H^2 (4\langle\eta\rangle^2 + \frac{1}{2}\langle\chi\rangle^2),
  ~~M^2_{\theta_3} = \frac{1}{2} g^2_{H} \langle\chi\rangle^2
\label{masses}
\end{equation}
The Yukawa couplings of the leptons are given by,
\begin{align}\nonumber
 {\cal L}_Y &= -h_1 \bar \psi_{eLi} \phi_ie_R - \tilde h_1\epsilon_{ij}\bar \psi_{eLi} \phi^j \nu_{eR}
 -h_2 \bar \Psi_{Li\alpha}\phi_i E_{R\alpha}-\tilde h_2\epsilon_{ij}\bar \Psi_{Li\alpha}\phi^j N_{R\alpha}\\
 &-k_2 \bar \Psi_{Li\alpha}\eta^\beta_{i\alpha} E_{R\beta}
 -\tilde k_2 \epsilon_{ij}\bar \Psi_{Li\alpha}\eta^{j\beta}_\alpha N_{R\beta} 
 -h_3 \bar \psi_{\tau Li} \phi_i \tau_R - \tilde h_3\epsilon_{ij}\bar \psi_{\tau Li} \phi^j \nu_{\tau R}+ \rm h.c 
\end{align}
which generate the following masses for leptons,
\begin{align}\nonumber
 m_e &= h_1 \langle\phi\rangle,
 ~~m_\tau = h_3 \langle\phi\rangle,
  ~~m_{\nu_e} = \tilde h_1 \langle\phi\rangle, 
 ~~m_{\nu_\tau} = \tilde h_3 \langle\phi\rangle\\
 ~~m_\mu &= h_2 \langle\phi\rangle + k_2 \langle\eta\rangle,
 ~~ m_{\nu_\mu} = \tilde h_2 \langle\phi\rangle + \tilde k_2 \langle\eta\rangle,\\
 m_{\mu^\prime} &= h_2 \langle\phi\rangle - k_2 \langle\eta\rangle,
 ~~m_{\nu_{\mu^\prime}} = \tilde h_2 \langle\phi\rangle - \tilde k_2 \langle\eta\rangle,\nonumber
 \label{mass}
 \end{align}
The required lepton masses can be generated by choosing the appropriate values of Yukawas.
\begin{table}
\caption{The Representation of the various fields in the model under the gauge group $G_{STD}\times SU(2)_{HV}$. Here $i$
and $\alpha$ are the $SU(2)_L$ and $SU(2)_{HV}$ indices respectively, which run through the value 1 and 2.}
\label{tab:1}       
%
%
\begin{tabular}{p{5cm}p{6.5cm}}
\hline\noalign{\smallskip}
Particles & $G_{STD}\times SU(2)_{HV} ~\rm quantum~numbers$  \\
\hline\noalign{\smallskip}
$\psi_{eLi} \equiv (\nu_e,e)$ & $(1,2,-1,1)$ \\
$\Psi_{Li\alpha} \equiv (\psi_\mu,\psi_{\mu^\prime})$ & $(1,2,-1,2)$\\
$\psi_{\tau Li} \equiv (\nu_\tau,\tau)$ & $(1,2,-1,1)$ \\
$E_{R\alpha} \equiv (\mu_R,\mu^\prime_R)$ & $(1,1,-2,2)$ \\
$N_{R\alpha} \equiv (\nu_{\mu R},\nu_{\mu^\prime R})$ & $(1,1,0,2)$ \\
$e_R,\tau_R$ & $(1,1,-2,1)$ \\
$\nu_{e R},\nu_{\tau R}$ & $(1,1,0,1)$\\
$\phi_i$ & $(1,2,1,1)$ \\
$\eta^\beta_{i\alpha}$ & $(1,2,1,3)$\\
$\chi_\alpha$ & $(1,1,0,2)$ \\
\noalign{\smallskip}\hline\noalign{\smallskip}
\end{tabular}
\end{table}
\section{Dark Matter Phenomenology}
\label{sec:2}
\begin{figure}[t]
\sidecaption[t]
\includegraphics[scale=0.80]{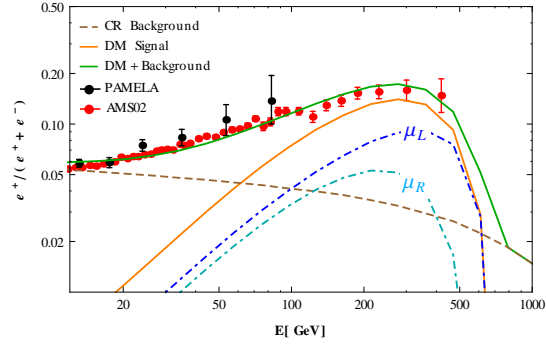}
\caption{The positron flux spectrum measured by AMS-02 \cite{Accardo:2014lma} and PAMELA \cite{Adriani:2008zr} (red and black
points respectively) and best fit using the contributions of different channels $(\mu_L$, $\mu_R)$ in our model.}
\label{fig:1}       
\end{figure}

In this model, the 4th generation right-handed neutral lepton $(\nu^\prime_{\mu_R}\equiv\chi)$ is identified as dark matter.
The only possible dark matter annihilation final states are $(\mu^+\mu^-)$ and $(\nu^c_\mu~\nu_\mu)$. 
To get the correct relic density \cite{6}, we use the Breit-Wigner resonant enhancement and take 
$M_{\theta_3} \simeq 2m_{\chi}$. By taking the 4th generation charged lepton $\mu^\prime$ heavier than $\chi$, the dark matter
stability is insured. The annihilation rate of $\chi$ for a single channel in the limit of massless leptons, is given as
\begin{equation}
 \sigma v = \frac{1}{16\pi} \frac{g^4_H m^2_\chi}{(s-M^2_{\theta_3})^2 + \Gamma^2_{\theta_3} M^2_{\theta_3}}
 \label{tcs}
\end{equation}
where $g_H$ is the horizontal gauge boson coupling, $m_\chi$ the dark matter mass, 
$M_{\theta_3}$ and $\Gamma_{\theta_3}$  are the mass and 
the decay width of $SU(2)_{HV}$ gauge boson respectively.
By calculating thermal average of annihilation cross-section rate using eq.(\ref{tcs}) and solving Boltzmann equation, 
we get the desired relic density of dark matter. The required parameters for getting the correct relic density is given in 
table.\ref{tab:2}.\\
\begin{table}
\caption{Numerical values of the parameters.}
\label{tab:2}
\begin{tabular}{p{1cm}p{1cm}p{1cm}p{1cm}p{1.7cm}p{1.7cm}p{1.7cm}p{1.7cm}} \\
\hline\noalign{\smallskip}
$g_H$  & $y_h$ & $y_A$ & $y_{H^\pm}$ & $m_{\chi}$ & $m_{\mu^\prime}$ & $M_{\theta_3,\theta^+}$ & $m_{H^\pm}$\\
\hline\noalign{\smallskip}
0.087 & 0.037 & 0.020 & 0.1 & 700 GeV& 740 GeV& 1400 GeV & 1700 GeV\\
\noalign{\smallskip}\hline\noalign{\smallskip}
\end{tabular}
\end{table}
In this model, the dark matter annihilates to $\mu^+\mu^-$ final state and their further decay produce positrons, which we use to explain
the positron excess seen at AMS-02. We use the publicly available code PPPC4DMID to compute the positron spectrum 
$\frac{dN_{e^+}}{dE}$ and then forward it to the GALPROP code for its propagation. To fit the AMS-02 data, the 
required cross-section (CS) in the GALPROP is $\sigma v_{\chi\chi\rightarrow \mu^+\mu^-}=2.3\times 10^{-25}\rm cm^3 s^{-1}$. 
The annihilation CS for $\mu$ final state from eq.(\ref{tcs}) is $\sigma v \approx 2.8 \times 10^{-25}\rm cm^3 s^{-1}$, which 
signify that in the fitting of AMS-02 data, we do not need astrophysical boost factor. In fig.(\ref{fig:1}), we compare the 
output of GALPROP code to the AMS-02 data and find it in well agreement with the data.
\section{Muon magnetic moment}
The $SU(2)_{HV}$ symmetry gives additional contributions to the muon $g-2$. The diagrams containing gauge boson $\theta^+$ and 
scalar $\eta^\beta_{i\alpha}$ cause extra contributions to muon $g-2$. 
In the limit of $M^2_{\theta^+}>>m^2_{\mu^\prime}$, the contribution from $SU(2)_{HV}$ gauge boson $\theta^+$ comes,
\begin{equation}
 [\Delta a_{\mu}]_{\theta^+} =  \frac{g^2_H}{8 \pi^2}\left(\frac{m_{\mu}m_{\mu^\prime}-2/3 m^2_{\mu}}{M^2_{\theta^+}}\right)
 \label{nmm}
\end{equation}
we note from the first term in eq.(\ref{nmm}) that there is $m_{\mu} m_{\mu^\prime}$ enhancement in the muon $(g-2)$. 
In the limits $m^2_{\mu^{\prime}}\gg m^2_h$, $m^2_{\mu^{\prime}}\gg m^2_A$, the contribution to muon $g-2$ from neutral higgs 
$\eta$ (CP-even $h$ and CP-odd $A$) comes,
\begin{equation}
 [\Delta a_\mu]_{h,A} = \frac{1}{8\pi^2}\left(\frac{3 m_\mu m_{\mu^\prime} (y^2_h-y^2_A) + m^2_\mu(y^2_h+y^2_A)}{6m^2_{\mu^\prime}}\right)
\end{equation}
where $y_h$ and $y_A$ are the Yukawa couplings of CP-even and CP-odd higgs respectively. In the similar way, the contribution 
from the charged higgs $\eta^\pm$ is given by,
\begin{equation}
 [\Delta a_\mu]_{H^{\pm}} = -\frac{y^2_{H^\pm}}{8\pi^2}\left(\frac{ 3 m_\mu m_{\nu_{\mu^\prime}}+m^2_\mu}{6m^2_{H^\pm}}\right)
\end{equation}
The total contribution to the muon anomalous magnetic moment is given as, 
\begin{equation}
 \Delta a_\mu = [\Delta a_\mu]_{\theta^+} +  [\Delta a_\mu]_{h,A} + [\Delta a_\mu]_{H^{\pm}}
\end{equation}
by taking into account the parameters shown in table.\ref{tab:2}, we finally get, 
\begin{equation}
 \Delta a_\mu = 2.9 \times 10^{-9}
\end{equation}
which is in agreement with the experimental result \cite{Bennett:2008dy} within $1\sigma$.
\section{Conclusion}
We studied a 4th generation extension of SM introducing $SU(2)_{HV}$ gauge symmetry between 4th generation fermions and muon families.
We identified the 4th generation neutral lepton as dark matter and proposed a common explanation for AMS-02 positron excess
and muon anomalous magnetic moment. The dark matter annihilates through $SU(2)_{HV}$ gauge boson $\theta_3$ and gives the 
correct relic density. The muons produced from the dark matter annihilation further decays and give positrons, which is used
for the explanation of AMS-02 positron excess. The $SU(2)_{HV}$ gauge boson $\theta^+$ and scalars give the 
additional contributions to the muon $g-2$. We found that for suitable choice of parameters, it is possible to get muon $g-2$
within $1\sigma$ of the BNL measurement.
\begin{acknowledgement}
I would like to thank the organizers of XXI DAE-BRNS High Energy Physics Symposium 2014 for inviting me and giving me 
opportunity to present my work. I also like to thank Prof. Subhendra Mohanty for valuable discussion. 
This contribution is based on our recent work \cite{Tomar:2014rya}.
\end{acknowledgement}

\end{document}